\documentclass[pdflatex,sn-mathphys-num]{sn-jnl}

\usepackage{graphicx}%
\usepackage{multirow}%
\usepackage{amsmath,amssymb,amsfonts}%
\usepackage{amsthm}%
\usepackage{bm}
\usepackage{mathrsfs}%
\usepackage[title]{appendix}%
\usepackage{xcolor}%
\usepackage{textcomp}%
\usepackage{manyfoot}%
\usepackage{booktabs}%
\usepackage{algorithm}%
\usepackage{algorithmicx}%
\usepackage{algpseudocode}%
\usepackage{listings}%
\usepackage{physics}
\usepackage{quantikz}
\usepackage{float}

\theoremstyle{thmstyleone}%
\theoremstyle{thmstyletwo}%

\theoremstyle{thmstylethree}%

\begin{document}

\title[Article Title]{Performance Comparison of QAOA Mixers for Ternary Portfolio Optimization}


\author*[1]{\fnm{Shintaro} \sur{Yamamura}}
\email{1225570@ed.tus.ac.jp}
\equalcont{These authors contributed equally to this work.}

\author[2]{\fnm{Satoshi} \sur{Watanabe}}
\email{xsas-watanabe@kddi.com}
\equalcont{These authors contributed equally to this work.}

\author[1]{\fnm{Masaya} \sur{Kunimi}}

\author[2]{\fnm{Kazuhiro} \sur{Saito}}

\author[1]{\fnm{Tetsuro} \sur{Nikuni}}

\affil*[1]{\orgdiv{Department of Physics},
  \orgname{Tokyo University of Science},
  \orgaddress{\street{1-3 Kagurazaka}, \city{Shinjuku},
  \postcode{162-8601}, \state{Tokyo}, \country{Japan}}}

\affil[2]{\orgname{KDDI Research, Inc.},
  \orgaddress{\street{2-1-15 Ohara}, \city{Fujimino},
  \state{Saitama}, \postcode{356-8502}, \country{Japan}}}


\abstract{The Quantum Approximate Optimization Algorithm (QAOA) is a quantum algorithm proposed for Noisy Intermediate-Scale Quantum (NISQ) devices and is regarded as a promising approach to combinatorial optimization problems, with potential applications in the financial sector. In this study, we apply QAOA to the portfolio optimization problem, which is one of the central challenges in financial engineering. A portfolio consists of a combination of multiple assets, and the portfolio optimization problem aims to determine the optimal asset allocation by balancing expected return and risk. In the context of quantum optimization, portfolio optimization is often formulated using discrete variables. Unlike conventional binary formulations, we consider a ternary portfolio optimization problem that accounts for three states—holding, not holding, and short selling—and compare its performance using different mixer operators. Specifically, we implement QAOA with the standard mixer and several XY Mixers (XY Ring, XY Parity Ring, XY Full, and QAMPA), and conducted simulations using real data based on the German stock index (DAX 30) for portfolios consisting of 5 and 8 assets. Furthermore, we introduce noise based on a depolarizing channel to investigate the behavior of the algorithm in realistic environments. The results show that while XY Mixers exhibit superiority in noiseless settings, their advantage degrades in noisy environments, and the optimal choice of mixer depends on both the number of QAOA depths and the noise strength.
}

\keywords{QAOA, Portfolio Optimization, QAOA Mixers, Financial Engineering, Quantum Computing, Noise Robustness}



\maketitle

\section{Introduction}

Quantum computing has been intensively studied as a new computational paradigm that exploits quantum mechanical principles to process information~\cite{Feynman1982}. Following the early conceptualization of quantum Turing machines~\cite{Deutsch1985}, several foundational algorithms were established. A landmark result is Shor's algorithm~\cite{Shor1994}, which demonstrated that integer factorization can be solved in polynomial time on a quantum computer, providing a concrete example of a potential exponential speedup over classical algorithms. Alongside this breakthrough, Grover's algorithm for unstructured search~\cite{Grover1996} and Lloyd's proposal for quantum simulation~\cite{Lloyd1996} were pivotal. Since then, various quantum algorithms have been proposed for problems such as optimization and simulation, establishing quantum computation as a promising direction in computational science.

Despite this theoretical progress, currently available quantum devices are limited in both qubit number and coherence time, and are commonly referred to as Noisy Intermediate-Scale Quantum (NISQ) devices~\cite{Preskill2018, Bharti2022}. In this regime, error-corrected quantum computation is not yet feasible, and practical quantum algorithms must be designed to operate under noise and hardware constraints. Variational quantum algorithms (VQAs), which combine parametrized quantum circuits with classical optimization, have therefore attracted considerable attention as candidates for near-term quantum applications~\cite{Cerezo2021, Preskill2018}. Prominent examples include the Variational Quantum Eigensolver (VQE) for quantum chemistry~\cite{Peruzzo2014, McClean2016, Tilly2022}, quantum machine learning (QML)~\cite{Benedetti2019, Mitarai2018, Schuld2019}, and quantum metrology~\cite{Koczor2020}. Among these, the Quantum Approximate Optimization Algorithm (QAOA)~\cite{Farhi2014}, originally proposed as a gate-based quantum algorithm for solving combinatorial optimization problems, is one of the most prominent VQAs. QAOA alternates between problem-dependent cost operators and problem-independent mixer operators, with variational parameters optimized classically.
Due to its shallow circuit structure and conceptual simplicity, QAOA has been widely investigated in the context of NISQ devices~\cite{Zhou2020, Wang2021, Harrigan2021}.

Portfolio optimization is a fundamental problem in financial engineering, aiming to determine an optimal allocation of assets by balancing expected return and risk~\cite{Markowitz1952, Campbell1997, Cochrane2009}.
The theoretical foundation of this problem was established by Markowitz through the mean-variance framework~\cite{Markowitz1952}, in which risk is quantified by the variance of portfolio returns.
While the original formulation leads to a convex quadratic optimization problem, practical portfolio optimization often involves additional constraints such as integer decision variables and cardinality constraints~\cite{Bienstock1996}.
It is well established that portfolio optimization problems incorporating such discrete constraints are computationally difficult.
In particular, mean-variance portfolio optimization with cardinality constraints has been shown to be NP-hard, implying that exact solution methods do not scale efficiently with problem size~\cite{Gao2013, Chang2000, Kellerer2000}.
To address this complexity, various classical heuristic approaches have been developed, including genetic algorithms~\cite{Chang2000, Soleimani2009}, simulated annealing~\cite{Crama1989}, tabu search~\cite{Glover1986, Woodside2011}, and particle swarm optimization~\cite{Cura2009, Zhu2011}.
While effective, these methods do not guarantee convergence to the global optimum, motivating the exploration of quantum approaches.

In recent years, quantum approaches to portfolio optimization have been actively explored~\cite{Orus2019, Egger2020}.
Beyond portfolio optimization, quantum algorithms are being applied to a wide range of financial problems, such as option pricing~\cite{Rebentrost2018, Stamatopoulos2020, Martin2021}, risk analysis~\cite{Woerner2019, Egger2020}, and amplitude estimation for Monte Carlo integration~\cite{Brassard2002, Suzuki2020}. Recent reviews discuss the broad potential of quantum algorithms in finance \cite{Bouland2020, Herman2023}.
Quantum annealing~\cite{Kadowaki1998} has been applied to financial optimization problems by mapping them to Ising or quadratic unconstrained binary optimization (QUBO) formulations, and experimental and numerical studies have demonstrated its applicability to realistic problem instances~\cite{Venturelli2019, Rosenberg2016}.
In parallel, gate-based quantum algorithms, particularly QAOA, have been proposed as an alternative framework for portfolio optimization on NISQ devices~\cite{Hodson2019, Brandhofer2022, Slate2021}.

Most existing studies of QAOA-based portfolio optimization adopt a binary representation, where each asset is described by a two-valued variable indicating whether the asset is held or not.
This formulation allows a straightforward mapping to qubit-based quantum circuits and has therefore been widely used in benchmarking and proof-of-concept studies~\cite{Brandhofer2022}.
However, real financial markets allow for more complex trading strategies, including the possibility of taking short positions.
From a modeling perspective, this motivates a ternary representation, in which each asset can take one of three states: long, no position, or short.
While such a representation provides a more faithful description of realistic portfolio construction, it introduces additional challenges for QAOA implementations.
In particular, the encoding of multi-valued variables, the preparation of feasible initial states, and the design of suitable mixer operators require careful consideration.
It has been shown that mixer design can significantly affect the performance of QAOA, especially in constrained optimization problems~\cite{Hadfield2019, WangXY2020}.
Nevertheless, to the best of our knowledge, a systematic comparison of different mixer operators for ternary portfolio optimization problems, including an assessment of their robustness under noise, has not yet been reported.

In this work, we implement QAOA for a ternary portfolio optimization problem and quantitatively evaluate the performance of several mixer operators using numerical simulations.
Specifically, we compare the Standard Mixer, XY Ring Mixer, XY Parity Ring Mixer, XY Full Mixer, and QAMPA (Quantum Alternating Mixer-Phase Ansatz).
By comparing these Mixers, we aim to clarify their relative effectiveness and provide design insights for applying QAOA to multi-valued financial optimization problems.
Furthermore, recognizing that noise is unavoidable in NISQ devices, we investigate how the relative performance of different Mixers is affected by noise, thereby identifying regimes in which specific mixer designs are preferable.

The structure of this paper is as follows: Section 2 formulates the portfolio optimization problem.
Section 3 explains the QAOA algorithm and the specific Mixers used in this study.
Section 4 describes classical optimization methods for tuning variational parameters.
Section 5 presents numerical simulation results and evaluates the effectiveness of the proposed approach.
Finally, Section 6 summarizes the conclusions of this study.
\section{Portfolio optimization}\label{sec2}
The portfolio optimization problem aims to determine the optimal portfolio weights by balancing expected return against the risk, quantified by the variance of returns. This problem is typically formulated as an optimization problem in which a cost function incorporating both risk and return is minimized.

In classical portfolio theory, the portfolio optimization problem is formulated based on Markowitz’s mean-variance model and solved as a continuous-variable optimization problem~\cite{Markowitz1952}. In contrast, for implementations on quantum devices, the problem is commonly formulated as a combinatorial optimization problem using discrete variables. 
In particular, ternary portfolio optimization models that explicitly incorporate short positions have been proposed~\cite{Hodson2019}. In this study, we adopt such a ternary formulation. We consider a portfolio consisting of $N$ assets, indexed by $i=1,\cdots,N$. Each asset is represented by a ternary variable $z_i \in \{1,0,-1\}$ corresponding to a long position, no position, or a short position, respectively. The cost function is defined as
\begin{align}
    F(z_1,\ldots,z_N) = q\sum_{i=1}^{N}\sum_{j=1}^{N}\sigma_{ij}z_iz_j - (1-q)\sum_{i=1}^{N}\mu_iz_i.\label{eq:cost_func}
\end{align} 
The first term represents the portfolio risk, while the second term corresponds to the expected portfolio return. 
Here, $\sigma_{ij}$ denotes the covariance between the returns of asset $i$ and $j$.
The parameter $\mu_i$ is the expected return of asset $i$, and $q \in [0,1]$ 
controls the trade-off between risk and return.
Risk-averse investors prefer $q\simeq 1$, while return-seeking investors prefer $q\simeq 0$. In this study, we set $q=1/3$.
The estimation procedures for the covariance matrix elements $\sigma_{ij}$ and the expected returns $\mu_i$ are described in Appendix~\ref{appendix:Estimate return and risk in a portfolio}.

To reflect practical investment constraints, we introduce a constraint on the total number of selected assets:
\begin{align}
    \sum_{i=1}^{N}z_i=B.\label{eq:constraint}
\end{align}
A portfolio satisfying this constraint is referred to as a ``feasible portfolio''.

The portfolio optimization problem in this study is therefore formulated as the following constrained combinatorial optimization problem:
\begin{align}
    \min_{z_1,\ldots,z_N} F(z_1,\ldots,z_N)\quad \text{s.t.}\; \sum_{i=1}^{N}z_i = B.
\end{align}

\begin{figure}[h]
\centering
\includegraphics[width=0.9\textwidth]{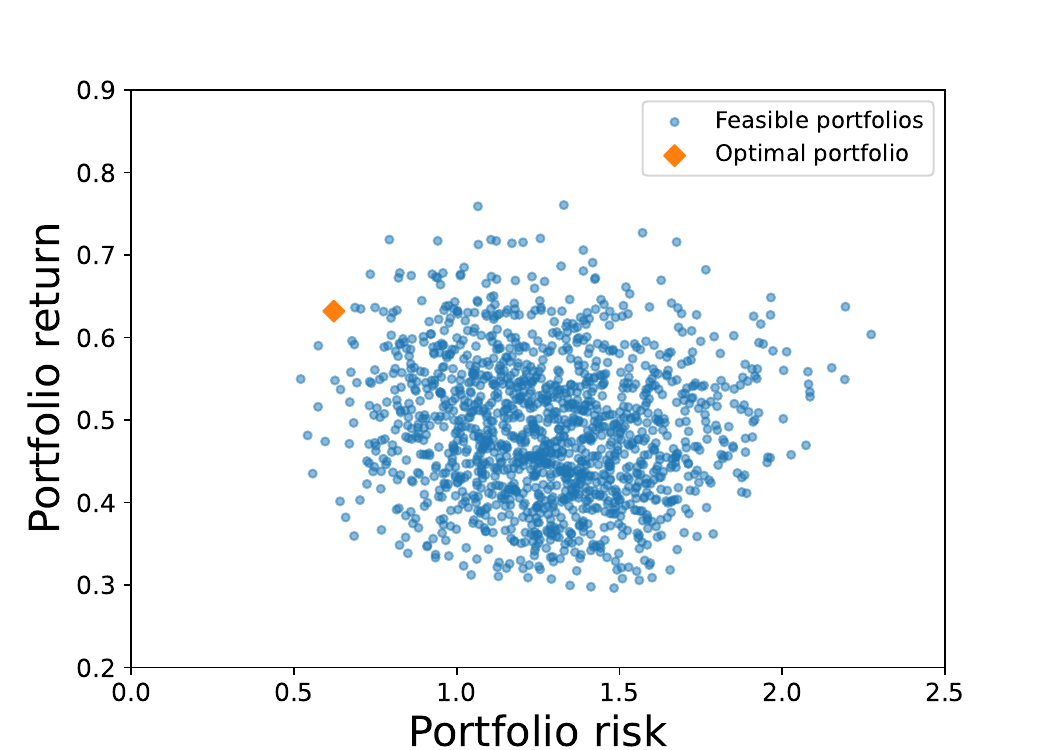}
\caption{
Feasible ternary portfolios in the risk-return plane for $N=8$ under the constraint $B = 4$.
Each feasible portfolio $z\in\{1,0,-1\}^N$ is shown as a circular marker, plotted to its portfolio risk $\sqrt{\sum_{i,j}\sigma_{ij} z_i z_j}$ on the horizontal axis and expected return $\sum_i \mu_i z_i$ on the vertical axis.
The triangular marker denotes the portfolio that minimizes the cost function $F(z)$ with $q=1/3$.
}
\label{fig:efficient_frontier}
\end{figure}

Figure.~\ref{fig:efficient_frontier} illustrates the set of feasible portfolios for $N=8$ under the constraint
$B=4$, plotted in terms of portfolio risk and expected return, using randomly generated expected returns and the covariance matrix.
Each circular marker corresponds to a feasible ternary portfolio, while the triangular marker highlights the portfolio that minimizes the cost function $F(z)$ defined above.
Accordingly, the orange point represents the optimal solution to the constrained combinatorial optimization problem.
The goal of portfolio optimization in this study is to efficiently identify this optimal portfolio among all feasible candidates.

\section{Evaluation metrics}\label{sec:evaluate_metrics}

This section introduces the metrics used to evaluate the performance of QAOA.
In this study, we adopt two metrics to quantitatively evaluate the obtained portfolios: the approximation ratio $r$ and the optimal solution probability $P$.

The approximation ratio $r$ is used as an index that measures how close the cost value of the obtained portfolio is to the optimal value among all feasible portfolios. It is defined as follows:
\begin{equation}
    r(z_1,\ldots,z_N) = 
    \begin{cases}
        \vspace{0.3em}\dfrac{F(z_1,\ldots,z_N)-F_{\max}}{F_{\min}-F_{\max}},\quad &\displaystyle{\sum_{i=1}^{N}z_i=B},\\
        0,&\displaystyle{\sum_{i=1}^{N}z_i\neq B},
    \end{cases}
\end{equation}
where
\begin{align}
    F_{\min} := \min_{\,\sum_{i}z_i=B}\; F(z_1,\ldots,z_N),\quad F_{\max} := \max_{\sum_{i}z_i=B}\; F(z_1,\ldots,z_N).
\end{align}
Here, $F_{\min}$ and $F_{\max}$ denote the minimum and maximum cost values, respectively, among feasible portfolios satisfying Eq.~\eqref{eq:constraint}.
In the simulation analysis, the average approximation ratio over multiple output distributions is used as a performance measure.

In addition, the optimal solution probability $P$ is defined as the probability that the measurement outcome corresponds to the optimal portfolio among all feasible portfolios.
It reflects how frequently QAOA outputs the optimal solution.

\section{Penalty method}

To handle constrained optimization problems using QAOA,
it is generally necessary to incorporate the constraints into the cost function when the quantum dynamics are not restricted to the feasible solution space. In such cases, the constraint is typically enforced by introducing penalty terms.

In this study, when the Standard Mixer is employed, the constraint on the number of selected assets is treated as a soft constraint by introducing a penalty term into the cost function, 
\begin{align}
    F^{(A)}(z_1,\ldots,z_N) = F(z_1,\ldots,z_N) + A\left(\sum_{i=1}^{N}z_i - B\right)^2.
\end{align}
Here, $A$ is a penalty coefficient designed such that the cost value increases for portfolios that violate the constraint.
By selecting a sufficiently large value of $A$, infeasible solutions are less likely to be chosen as optimal during the optimization process.

In contrast, when using the XY Mixer, it is theoretically possible to explore only the feasible solution space that satisfies the constraint. In this case, the constraint is preserved exactly throughout the QAOA evolution, effectively acting as a hard constraint. As a result, the introduction of a penalty term is unnecessary when the XY Mixer is used.

Accordingly, in this study, the penalty term is introduced only when the Standard Mixer is employed. The value of the penalty coefficient $A$ is chosen such that the average cost value of infeasible solutions is greater than or equal to that of feasible ones, ensuring that infeasible solutions are not favored during optimization~\cite{Brandhofer2022} (see  Alg.~\ref{alg:estimateA}).

\begin{algorithm}[t]
    \caption{Penalty efficient estimation for constraint enforcement}
    \label{alg:estimateA}
    \begin{algorithmic}[1]
        \Require $F^{(A)}(z)$ with penalty coefficient $A\ge 0$,
        constraint $g(z)=\sum_{i=1}^{N} z_i - B$,
        increment rule $\Delta A$.
        \Ensure Penalty coefficient $A$ such that the infeasible minimum is sufficiently separated
        \State $A \gets 0$
        \Repeat
            \State Compute the feasible minimum:
        \[
        F_{\min} \gets \min_{z:\, g(z)=0} F^{(A)}(z)
        \]
        \State Compute the feasible mean:
        \[
        \bar{F} \gets \frac{1}{\binom{2N}{N+B}} \sum_{z:\, g(z)=0} F^{(A)}(z)
        \]
        \State Compute the infeasible minimum and an argmin:
        \[
        (F_{\min}^{(\mathrm{inf})}, z^\star) \gets
        \arg\min_{z:\, g(z)\neq 0} F^{(A)}(z)
        \]
        \State Check separation condition:
        \[
        \text{if } F_{\min}^{(\mathrm{inf})} \ge \tfrac{1}{2}\bigl(F_{\min}+\bar{F}\bigr)
        \text{ then return } A
        \]
        \State Set increment using the current worst infeasible state $z^\star$:
        \[
        \Delta A \gets \frac{\tfrac{1}{2}\bigl(F_{\min}+\overline{F}\bigr) - F_{\min}^{(\mathrm{inf})}}
                       {g(z^\star)^2}
        \]
        \State $A \gets A + \Delta A$
        \Until{condition is satisfied}
    \end{algorithmic}
\end{algorithm}
\section{QAOA algorithm}\label{sec5}

The QAOA is a variational algorithm based on quantum gate operations, designed to obtain approximate solutions to combinatorial optimization problems.

Starting from an initial state $|\psi_0\rangle_M$, two types of parameterized unitary operators, the cost unitary $\hat{U}_F(\gamma)$ and the mixer unitary $\hat{U}_M(\beta)$, are alternately applied $p$ times.
Here, $\mathrm{M}$ specifies the type of mixer hamiltonian.
The resulting final state $|\psi_{\vec{\gamma},\vec{\beta}}\rangle_M$
is constructed as follows:
\begin{align}
    |\psi_{\vec{\gamma},\vec{\beta}}\rangle_M = \prod_{k=1}^{p}\hat{U}_M(\beta_k)\hat{U}_F(\gamma_k)|\psi_0\rangle_M.
\end{align}
Here, $\hat{U}_M(\gamma_k)$ and $\hat{U}_F(\beta_k)$ are respectively defined as
\begin{align}
    \hat{U}_M(\beta_k) = e^{-i\beta_k\hat{M}}, \quad \hat{U}_F(\gamma_k)=e^{-i\gamma_k\hat{F}}.
\end{align}
The cost Hamiltonian $\hat{F}$ represents the cost function, expressed as an operator obtained by replacing the classical variables $z_1,\ldots,z_N$ with their corresponding quantum operators. The mixer Hamiltonian $\hat{M}$ plays the role of exploring the solution space by inducing transitions between computational basis states.

In general, when the mixer consists of non-commuting terms, the exact time-evolution operator generated by a single mixer Hamiltonian cannot be implemented directly. In QAOA, the mixer is therefore specified at the level of unitary ansatz, defined as a product of elementary unitaries. Throughout this work, we present the mixer directly in terms of the corresponding unitary operators, without explicitly introducing a single underlying mixer Hamiltonian.

In QAOA, the variational parameters
\begin{align}
    \vec{\gamma}=(\gamma_1,\ldots,\gamma_p)\in [0,\pi]^p, \quad \vec{\beta} = (\beta_1,\ldots,\beta_p)\in[0,\infty)^p,
\end{align}
are classically optimized so as to minimize the expectation value of the cost Hamiltonian:
\begin{align}
    \langle \hat{F} \rangle_{\vec{\gamma},\vec{\beta}} = \langle \psi_{\vec{\gamma},\vec{\beta}} |\hat{F}|\psi_{\vec{\gamma},\vec{\beta}}\rangle,\label{eq:exp}
\end{align}
thereby yielding an approximate optimal solution.

While this study focuses on the standard QAOA framework with various Mixers, several advanced variants have been proposed to enhance performance.
These include Recursive QAOA (R-QAOA) which iteratively eliminates variables, Warm-start QAOA initialized with classical solutions \cite{Egger2021, Tate2023}, and adaptive strategies (ADAPT-QAOA) that dynamically construct the ansatz \cite{Zhu2022, Wurtz2021}.
Furthermore, addressing the trainability issues caused by barren plateaus in the optimization landscape remains a critical challenge for scaling these algorithms \cite{McClean2018, Wang2021, Holmes2021, Arrasmith2021}.
\subsection{Encoding and cost Hamiltonian}\label{subsec:encording and cost hamiltonian}
To implement the optimization problem on a quantum computer, the cost function $F^{(A)}$ must be converted into a corresponding cost Hamiltonian.
In this study, we encode the asset-holding state of each asset using qubits.
Following the encoding scheme proposed in previous work~\cite{Hodson2019}, we represent the ternary asset-holding state using two qubits per asset.

Specifically, for asset $i$, we introduce binary variables $x_i^+,x_i^-\in \left\{0,1\right\}$, and represent the selection variable $z_i$ as
\begin{equation}
    z_i = x_i^+ - x_i^-.\label{eq:z_encoding}
\end{equation}
Here, $x_i^+=1$ indicates that the asset is held, while $x_i^-=1$ indicates that the asset is sold.
By associating these binary variables with quantum states, the holding state of each asset is encoded by the two-qubit state $|x_i^-,x_i^+\rangle$.

In this encoding, the three possible portfolio positions are represented as follows:
the long position corresponds to $(x_i^{-},\,x_i^{+})=(0,1)$, the short position to $(1,0)$, and the no-position state to $(0,0)$ and $(1,1)$. This encoding introduces a degeneracy in the no-position state, since both $(0,0)$ and $(1,1)$ correspond to $z_i=0$; however, this degeneracy does not affect the cost function or the constraint, which depend only on the aggregated variable defined in Eq.~\eqref{eq:z_encoding}.

Consequently, an overall portfolio configuration consisting of $N$ assets is represented by a $2N$-qubit state as
\begin{equation}
    (z_1,\ldots,z_N)\to|x_1^-,x_1^+,\ldots, x_N^-,x_N^+\rangle.
\end{equation}
Hereafter, the qubit corresponding to $x_i^-$ is referred to as the “short qubit”, and the qubit corresponding to $x_i^+$ as the ``long qubit".

Since $z_i$ can be mapped to 
\begin{equation}
    z_i\to \frac{\hat{Z}_i^--\hat{Z}_i^+}{2},
\end{equation}
by introducing the Pauli-$Z$ operators $\hat{Z}_i^-$ and $\hat{Z}_i^+$ acting on the short and long qubits, respectively, the cost function can be transformed into the following operator form:
\begin{align}
    \hat{F} = \lambda F^{(A)}\left(\frac{\hat{Z}_1^--\hat{Z}_1^+}{2},\ldots, \frac{\hat{Z}_N^--\hat{Z}_N^+}{2}\right),
\end{align}
where $\lambda$ is a scaling factor introduced to improve the numerical stability of parameter optimization in QAOA. For the Standard Mixer, $\lambda$ is given by $4N$, while for the XY Mixers and QAMPA it is given by $2N(2N-1)$, reflecting the different numbers of terms contributing to the mixer Hamiltonian~\cite{Brandhofer2022}.

Accordingly, the cost unitary $\hat{U}_F(\gamma)$ is given by
\begin{align}
    \hat{U}_F(\gamma)=
    \prod_{i<j}
    e^{-i\gamma W_{ij}(\hat{Z}_i^{-}\hat{Z}_j^{-}
    - \hat{Z}_i^{-}\hat{Z}_j^{+}
    - \hat{Z}_i^{+}\hat{Z}_j^{-}
    + \hat{Z}_i^{+}\hat{Z}_j^{+})}
    \prod_{i=1}^{N}
    e^{i\gamma w_{i}(\hat{Z}_{i}^{+}-\hat{Z}_{i}^{-})},\label{eq:cost_hamiltonian}
\end{align}
where the coefficients $W_{ij}$ and $w_i$ are defined as follows:
\begin{align}
    W_{ij} = \frac{\lambda}{2} (q\sigma_{ij} + A), \quad 
    w_{i} = \frac{\lambda}{2}\left[(1-q)\mu_i + 2AB\right].
\end{align}
\subsection{Mixer unitary}\label{subsec:Mixer_unitary}

The mixer unitary plays a role in exploring quantum states. In this study, we conduct a comparative analysis of five different mixer unitaries.

In this section, we specify the Mixers directly through their unitary operators used in the QAOA circuit. Although these unitaries can be formally associated with effective mixer Hamiltonians, we do not introduce explicit Hamiltonian expressions, since the mixer is implemented as a product of elementary unitaries constituting the variational ansatz. The design of the mixer Hamiltonian is crucial for constraining the search space and incorporating problem symmetries \cite{Hadfield2019, WangXY2020, Shaydulin2019}. Alternative approaches, such as Grover-Mixers \cite{Bartschi2020} or subspace-search mechanisms \cite{Cook2020, Marsh2019, Fuchs2022}, have also been proposed for constrained optimization.

\subsubsection{Standard Mixer (${M}=\mathrm{standard}$)}
The Standard Mixer~\cite{Farhi2014} is a mixer commonly used in QAOA.
It consists of Pauli-$X$ operators acting on all qubits and is defined as follows:
\begin{equation}
\hat{U}_{\mathrm{standard}}(\beta)
\equiv
\prod_{i=1}^{2N} e^{i\beta \hat{X}_i}.
\end{equation}
In this case, the initial state is set to the uniform superposition of all computational basis states:
\begin{equation}
    |\psi_0\rangle_{\text{standard}} = |+\rangle^{\otimes 2N},
\end{equation}
where $|+\rangle = (|0\rangle + |1\rangle)/\sqrt{2}$.
\subsubsection{XY Mixer ($M=M_{XY}$)}
The XY Mixer~\cite{Hadfield2019} is a mixer that enables exploration within the solution space that satisfies the constraints.
The initial state is prepared as a superposition of states satisfying the constraint:
\begin{equation}
|\psi_0\rangle_{{M}_{XY}} = \frac{1}{\sqrt{K}}
\sum_{\substack{x_1^-,\ldots,x_N^-,\,x_1^+,\ldots,x_N^+=0,1 \\ 
(x_1^+,\ldots,x_N^+)-(x_1^-,\ldots,x_N^-)=B}}
|x_1^- x_1^+ \cdots x_N^- x_N^+\rangle.\label{eq:XY_initial}
\end{equation}
Here, $K$ is a normalization constant defined as
\begin{align}
    K = \sum_{i=0}^{N-B}\binom{N}{B+i}\binom{N}{i}
    = \binom{2N}{N+B}.
\end{align}
Details of the circuit implementation for preparing the initial state $|\psi_0\rangle_{M_{XY}}$ are provided in Appendix~\ref{appendix:dicke}.

The two-qubit XY-type unitary acting on the $i$-th and 
$j$-th qubits is defined as
\begin{equation}
    \hat{R}_{i,j}^{(XY)}(\beta) = e^{i\beta\left(\hat{X}_i\hat{X}_j+\hat{Y}_i\hat{Y}_j\right)}. 
\end{equation}
This operator generates a coherent mixing between the states $|01\rangle$ and $|10\rangle$ of the qubit pair ($i,j$), while leaving $|00\rangle$ and $|11\rangle$ unchanged. The XY mixer unitary is then constructed as the product of these two-qubit operators, as given by
\begin{equation}
    \hat{U}_{M_{XY}}(\beta) = \prod_{(i,j)\in S_{M_{XY}}}\hat{R}_{i,j}^{(XY)}(\beta)
    = \prod_{(i,j)\in S_{M_{XY}}}e^{i\beta\left(\hat{X}_i\hat{X}_j+\hat{Y}_i\hat{Y}_j\right)}.
\end{equation}
Here, $S_{M_{XY}}$ denotes the set of qubit pairs ($i,j$) on which the XY mixer acts. In this study, we implement four types of XY Mixers as described below.
\subsubsection{XY Ring Mixer}
We consider two qubit strings, corresponding to the short and long qubits.  
We denote the index set of these strings by $\mathcal{L}=\{-,+\}$.
For each $\ell\in\mathcal{L}$, the qubit string is written as
\begin{equation}
    \boldsymbol{x}^{(\ell)}
      = (x^{(\ell)}_1, x^{(\ell)}_2, \ldots, x^{(\ell)}_{N}).
\end{equation}
For each string $\ell$, the set of adjacent qubit pairs with periodic
boundary conditions is defined as
\begin{equation}
    S_{\mathrm{ring}}^{(\ell)}
      = \{(x^{(\ell)}_1,x^{(\ell)}_2),(x^{(\ell)}_2,x^{(\ell)}_3)\ldots,(x^{(\ell)}_{N},x^{(\ell)}_1)\}.
\end{equation}
The global set is given by
\begin{equation}
    S_{\mathrm{ring}}=\bigcup_{\ell\in\mathcal{L}}S_{\mathrm{ring}}^{(\ell)}.
\end{equation}
The XY mixer unitary is then expressed as
\begin{align}
    \hat{U}_{M_{XY}}(\beta)
      = \prod_{\ell\in\mathcal{L}}
        \prod_{(i,j)\in S_{\mathrm{ring}}^{(\ell)}}
            \hat{R}_{i,j}^{(XY)}(\beta).
\end{align}

\subsubsection{XY Parity Ring Mixer}

For each $\ell\in\mathcal{L}$, we decompose the ring structure 
$S_{\mathrm{ring}}^{(\ell)}$ into two subsets according to the parity 
of the site index:
\begin{align}
    S_{\mathrm{par,even}}^{(\ell)}
      &= \{(x_i^{(\ell)}, x_{i+1\bmod N}^{(\ell)})
          \mid i=1,2,\ldots,N,\ i\equiv 0\ (\mathrm{mod}\ 2)\},\\
    S_{\mathrm{par,odd}}^{(\ell)}
      &= \{(x_i^{(\ell)}, x_{i+1\bmod N}^{(\ell)})
          \mid i=1,2,\ldots,N,\ i\equiv 1\ (\mathrm{mod}\ 2)\}.
\end{align}

These two subsets are disjoint, and their union reconstructs the original
ring structure, i.e.,
\begin{equation}
    S_{\mathrm{ring}}^{(\ell)}
      = S_{\mathrm{par,even}}^{(\ell)} \cup S_{\mathrm{par,odd}}^{(\ell)}.
\end{equation}

The corresponding XY Parity Ring Mixer is defined as
\begin{equation}
    \hat{U}^{\mathrm{par}}_{M_{XY}}(\beta)
      = \prod_{k\in\{\mathrm{even},\mathrm{odd}\}}
        \prod_{\ell\in\mathcal{L}}
        \prod_{(i,j)\in S_{\mathrm{par},k}^{(\ell)}}
            \hat{R}_{i,j}^{(XY)}(\beta),
\end{equation}
which allows the two layers (even and odd) to be executed in parallel,
thereby reducing the circuit depth compared with the XY Ring Mixer~\cite{Brandhofer2022}.
\subsubsection{XY Full Mixer}
The XY Full Mixer acts on all possible qubit pairs within each string
$\ell\in\mathcal{L}$, not only on adjacent ones~\cite{Brandhofer2022}.
For each qubit string $\boldsymbol{x}^{(\ell)} =
(x^{(\ell)}_1,x^{(\ell)}_2,\ldots,x^{(\ell)}_{N})$, we define
\begin{equation}
    S_{\mathrm{full}}^{(\ell)}
      = \{ (x_i^{(\ell)}, x_j^{(\ell)}) \mid
           1\le i < j \le N \},
\end{equation}
which contains all ${N\choose 2}$ possible pairs.

To reduce the circuit depth, the elements of $S_{\mathrm{full}}^{(\ell)}$ can be
partitioned into several disjoint layers such that no two pairs in the same
layer share a qubit, thereby enabling the parallel application of XY gates.

As an example, when $N=5$ and $\ell=-$, one possible decomposition of
$S_{\mathrm{full}}^{(-)}$ is
\begin{align}
S_{\mathrm{full}}^{(-)}
=
\{&(x_1^-,x_5^-), (x_2^-,x_4^-), (x_1^-,x_3^-),
  (x_3^-,x_5^-), (x_4^-,x_5^-), \notag \\
 &(x_1^-,x_2^-), (x_2^-,x_3^-), (x_3^-,x_4^-),
  (x_1^-,x_4^-), (x_2^-,x_5^-)\},
\end{align}
and the same construction applies to $\ell=+$ case.

The corresponding QAOA layer using the XY Full Mixer is written as
\begin{align}
    \hat{U}_{\mathrm{full}}(\beta)e^{-i\gamma\hat{F}}
     &=\prod_{\ell\in\mathcal{L}}
       \prod_{(i,j)\in S_{\mathrm{full}}^{(\ell)}}
           \hat{R}^{(\mathrm{XY})}_{i,j}(\beta) \nonumber\\[4pt]
     &\times
     \prod_{i<j} e^{-i\gamma W_{ij}
         (\hat{Z}_i^{-}\hat{Z}_j^{-}
          -\hat{Z}_i^{-}\hat{Z}_j^{+}
          -\hat{Z}_i^{+}\hat{Z}_j^{-}
          +\hat{Z}_i^{+}\hat{Z}_j^{+})} \nonumber\\[4pt]
     &\times
     \prod_{i=1}^{N}
        e^{\, i\gamma w_{i}
        (\hat{Z}_{i}^{+}-\hat{Z}_{i}^{-})}.
\label{eq:Full}
\end{align}

\subsubsection{QAMPA}

The QAMPA integrates the XY Full
Mixer and the cost unitary into a single unitary operation~\cite{Brandhofer2022}.
By avoiding controlled-NOT (CNOT) gates, this ansatz reduces circuit
depth and improves ~\cite{LaRose2022}.  
In particular, the total number of CNOT gates can be reduced to
approximately three-quarters of that used in the conventional XY Full Mixer.

Using the full interaction set
\begin{align}
S_{\mathrm{full}}^{(\ell)}
=\{(x_i^{(\ell)},x_j^{(\ell)})\mid 1\le i<j\le N\},
\quad \ell\in\mathcal{L},
\end{align}
the QAMPA mixer–phase unitary can be written as
\begin{equation}
\begin{aligned}
\hat{U}_{\mathrm{MP}}(\beta,\gamma)
 ={}&
\prod_{\ell\in\mathcal{L}}
\prod_{(i,j)\in S_{\mathrm{full}}^{(\ell)}}
   e^{\, i\beta(\hat{X}_i^{(\ell)}\hat{X}_j^{(\ell)}
               +\hat{Y}_i^{(\ell)}\hat{Y}_j^{(\ell)})
     - i\gamma W_{ij}\hat{Z}_i^{(\ell)}\hat{Z}_j^{(\ell)}} \\
&\times
\prod_{i<j}
   e^{\, i\gamma W_{ij}
      (\hat{Z}_i^{-}\hat{Z}_j^{+}
       +\hat{Z}_i^{+}\hat{Z}_j^{-})} \\
&\times
\prod_{i=1}^{N}
   e^{\, i\gamma W_{ii}
      \hat{Z}_{i}^{-}\hat{Z}_{i}^{+}} \\
&\times
\prod_{i=1}^{N}
   e^{\, i\gamma w_i
      (\hat{Z}_{i}^{-}-\hat{Z}_{i}^{+})}.
\end{aligned}
\label{eq:QAMPA}
\end{equation}

Accordingly, one step of QAOA with $p$ layers is expressed as
\begin{equation}
|\psi_{\vec{\gamma},\vec{\beta}}\rangle_{M_{XY}}
=
\prod_{k=1}^{p}
\hat{U}_{\mathrm{MP}}(\beta_k,\gamma_k)
\,|\psi_0\rangle_{M_{\mathrm{XY}}},
\end{equation}
where $|\psi_0\rangle_{M_{\mathrm{XY}}}$ denotes the initial state defined in Eq.~\eqref{eq:XY_initial}.

Details of the quantum circuit implementations for the QAOA construction are provided in Appendix~\ref{appendix:Quantum Circuits}.


\section{Parameter Optimization}
In QAOA, the variational parameters ($\gamma,\beta$) must be classically optimized to minimize the expectation value of the cost Hamiltonian defined as Eq.~\eqref{eq:exp}.
This optimization method is a key factor that significantly affects the performance of  QAOA. This section describes the classical parameter optimization used in this research.
\subsection{Initial parameter values for $p = 1$}
For QAOA depth $p=1$, appropriately setting the initial parameters ($\gamma,\beta$) can improve the convergence of the optimization. In this study, we employed a grid search to estimate the initial parameters for the case $p=1$.

First, using $m_1,m_2>0$, we define $\vec{\gamma}^{(\text{lin},p)}(m_1)$ and $\vec{\beta}^{(\text{lin},p)}(m_2)$ as follows:
\begin{align}
    &\vec{\gamma}^{(\text{lin},p)}(m_1) = m_1\left(x_1^{(p)},x_2^{(p)},\ldots,x_p^{(p)}\right),\\
    &\vec{\beta}^{(\text{lin},p)}(m_2) = m_2\left(1-x_1^{(p)},1-x_2^{(p)},\ldots,1-x_p^{(p)}\right).
\end{align}
Here, $x_i^{(p)}$ is defined as
\begin{equation}
    x_i^{(p)}=\frac{2i-1}{2p}.
\end{equation}

\begin{figure}[t]
\centering
\includegraphics[width=1.0\textwidth]{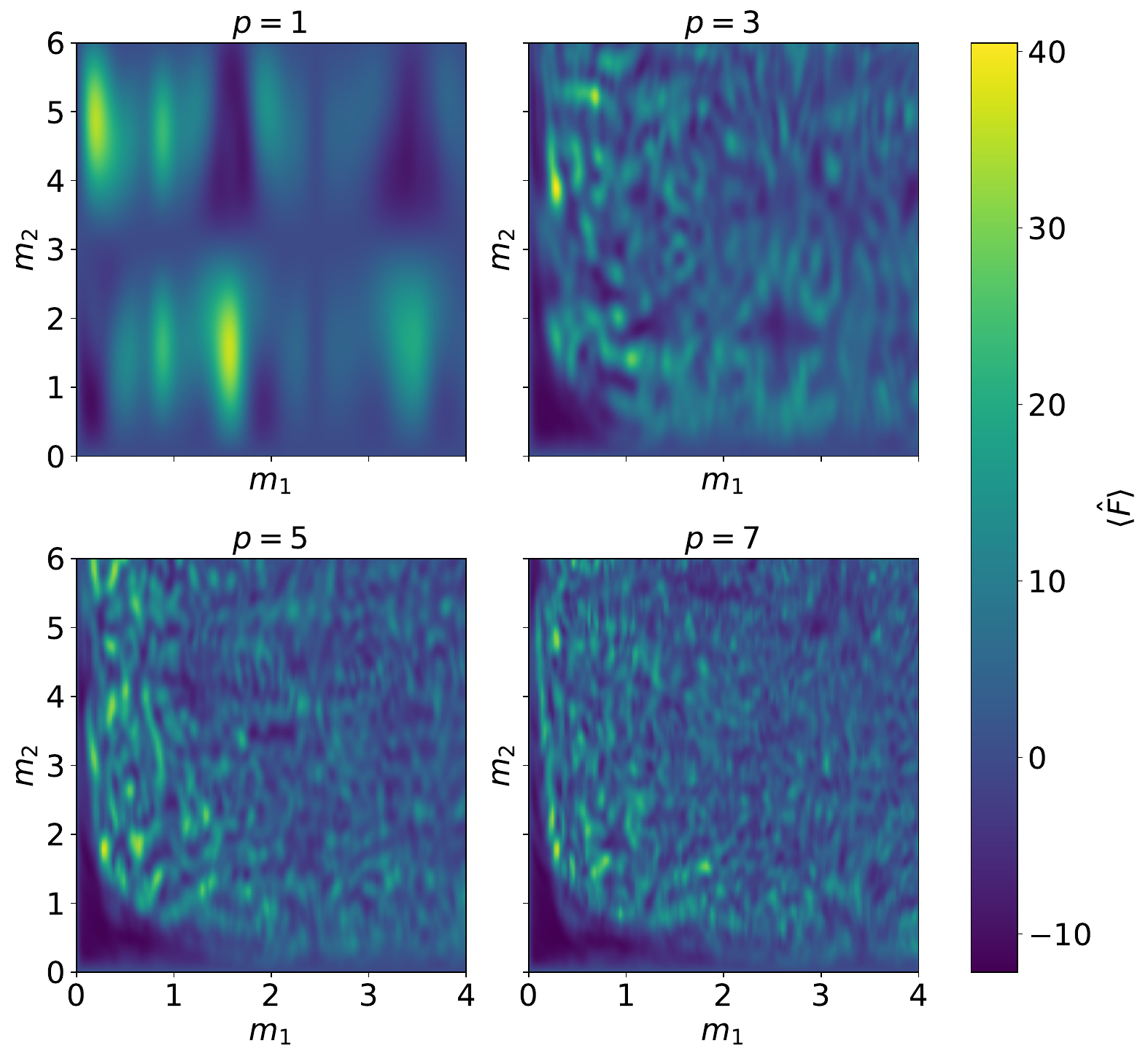}
\caption{Computed energy landscapes for depth parameters $p=1,3,5$, and $7$ in a portfolio with $n=5$ assets under the investment constraint $B=2$.}\label{fig:landscape}
\end{figure}

To illustrate the structure of the optimization landscape and to motivate the choice of initial parameters, we evaluated the expectation value

\begin{align}
    \langle \hat{F}\rangle_{m_1,m_2}^{(\text{lin},p)} = \langle \hat{F}\rangle_{\vec{\gamma}^{(\text{lin},p)}(m_1),\vec{\beta}^{(\text{lin},p)}(m_2)},
\end{align}
on the $m_1-m_2$ plane. Figure~\ref{fig:landscape} shows the resulting energy landscapes for $p=1,3,5$ and $7$, demonstrating how the landscape becomes increasingly structured as the QAOA depth increases.

\subsection{Initial parameter values for $p>1$}

It has been shown that the initial parameters for layer $p$ in QAOA can be extracted from the optimal parameters of the layer $p-1$ \cite{Zhou2020}.
In this study, four estimation methods based on prior study \cite{Brandhofer2022} were implemented, and the one for which Eq.~\eqref{eq:exp} is minimized was adopted.

\subsection{Classical optimization algorithms}

In this study, we simulate QAOA circuits using the Statevector Simulator and the Qasm Simulator implemented in Qiskit \cite{Qiskit}.
The Statevector Simulator computes quantum states exactly without any statistical sampling error.
In contrast, the Qasm Simulator performs probabilistic simulations based on a finite number of measurements (shots) and therefore introduces sampling noise.
For the classical optimization routine, we employ the gradient-based SLSQP method \cite{Kraft1988} with the Statevector Simulator, and the gradient-free Nelder–Mead method \cite{Nelder1965} with the Qasm Simulator.

\section{Results}
\subsection{$N=5$ with $B=2$}

\begin{figure}[h]
\centering
\includegraphics[width=1.0\textwidth]{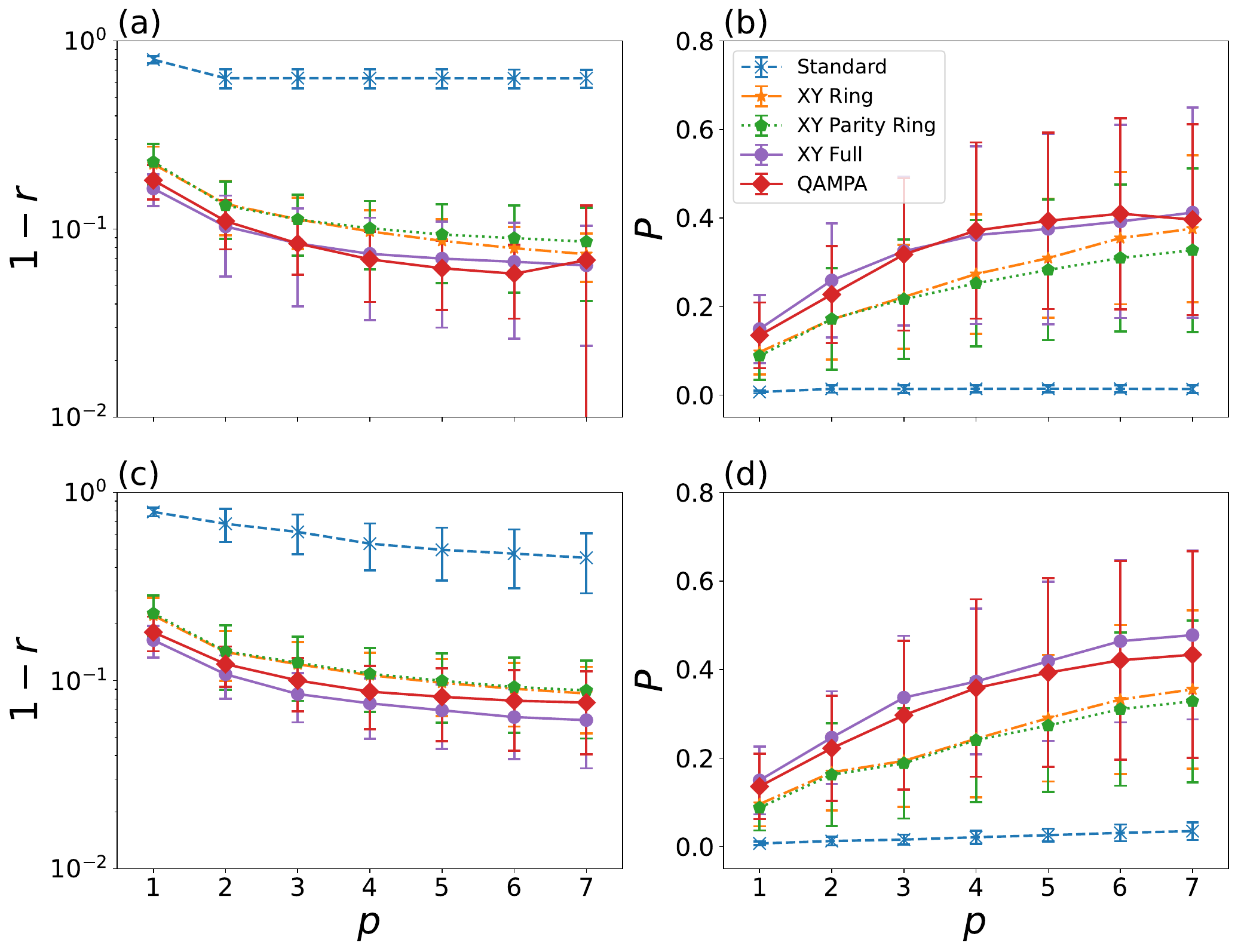}
\caption{Average approximation ratio and ground state probability for QAOA with five
diﬀerent Mixers, evaluated over 20 randomly generated portfolios with $N=5$ assets and investment constraint $B=2$, as functions of the QAOA depth $p$. Error
bars indicate standard deviations. The top row shows results obtained using the
Statevector Simulator: (a) average approximation ratio $1-r$, and (b) ground state probability $P$. The bottom row shows the results from the Qasm Simulator: (c) $1-r$, and (d) $P$.}\label{fig2}
\end{figure}

In this section, we present simulation results of QAOA using five Mixers and provide a comparative analysis. As evaluation metrics, we use the average approximation ratio $r$ and the optimal solution probability $P$, which were defined in Sec.~\ref{sec:evaluate_metrics}, as functions of the QAOA depth $p$.
For the simulations, we generated 20 random portfolios based on the DAX30 index under the conditions of asset count $N=5$ with investment constraint $B=2$, and $N=8$ with $B=4$.
In this subsection, we present the results for $N=5$ and $B=2$, obtained using both the Statevector Simulator and the Qasm Simulator, as illustrated in Fig.~\ref{fig2}.

Figures 2(a) and 2(b) show the results obtained using the Statevector Simulator. Even at a shallow QAOA depth of $p=1$, all XY Mixers outperform the Standard Mixer.
This is because the XY Mixers operate strictly within the subspace of feasible solutions that satisfy the constraint~\cite{Cook2020}.
Specifically, while the Standard Mixer explores all $2^{10}=1024$ candidate states, the XY Mixers restrict the search to only 120 feasible portfolios.
Among the XY Mixers, the XY Full Mixer and QAMPA converge most rapidly to the optimal solution, achieving a high approximation ratio of $r>99\%$.
This improvement arises because the XY Ring and XY Parity Ring Mixers couple only neighboring qubits, whereas the XY Full Mixer and QAMPA couple all qubit pairs, enabling broader transitions within a single layer. A similar trend is observed for the optimal solution probability $P$, with the XY Full and QAMPA reaching $P\simeq 0.6$.

On the other hand, Figs.~2(c) and 2(d) show the results obtained using the Qasm Simulator, where measurement noise leads to slower convergence toward the optimal solution compared to the Statevector Simulator. The number of measurement shots was set to 3000 for these simulations.

\subsection{$N=8$ with $B=4$}

\begin{figure}[h]
\centering
\includegraphics[width=1.0\textwidth]{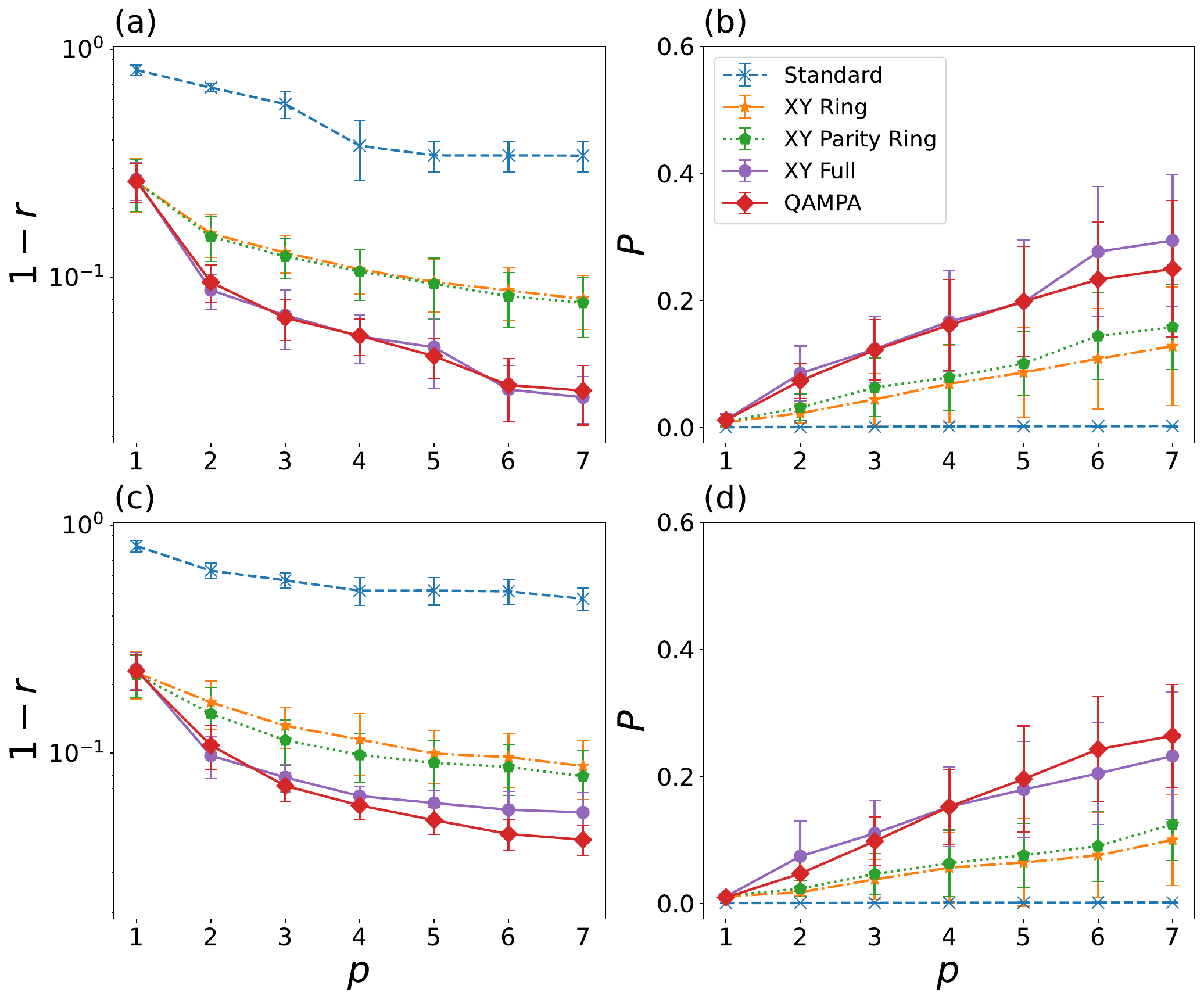}
\caption{Average approximation ratio and ground state probability for QAOA with five
diﬀerent Mixers, evaluated over 10 randomly generated portfolios with $N=8$ assets and investment constraint $B=4$, as functions of the QAOA depth $p$. Error
bars indicate standard deviations. The top row shows results obtained using the
Statevector Simulator: (a) average approximation ratio $1-r$, and (b) ground state probability $P$. The bottom row shows the results from the Qasm Simulator: (c) $1-r$, and (d) $P$.}\label{N=8}
\end{figure}

Next, Fig.~\ref{N=8} presents the results for the case expanded to $N=8$ assets with an investment constraint of $B=4$.
The number of shots in the Qasm Simulator was set to 10,000.
In this setting as well, the XY Full Mixer and the QAMPA Mixer exhibited superior performance compared with the other Mixers.
However, relative to the results of $1-r$ and $P$ shown in Fig.~\ref{fig2}, a degradation in overall performance was observed across all metrics.
A plausible explanation is that when the number of assets increases to $N=8$, the number of feasible portfolios grows to 15,504, thereby enlarging the search space and making the optimization problem more challenging.

\subsection{Performance under depolarizing Noise}

To evaluate the performance of QAOA in the presence of noise, we employed a depolarizing channel~\cite{Nielsen2010}. For a $2N$-qubit density matrix $\hat{\rho}$, the depolarizing channel is defined as
\begin{align}
    \mathcal{E}(\hat{\rho}) = \eta \frac{\hat{I}}{2^{2N}} + (1 - \eta)\hat{\rho},
\end{align}
where $\eta$ represents the noise strength and $\hat{I}$ is the $2^{2N}$-dimensional identity operator. In the QAOA circuit, the depolarizing channel is applied after every one-qubit and two-qubit gate to each target qubit.

We evaluated the performance of QAOA under various conditions by varying the QAOA depth $p$, the mixer type, and the noise strength $\eta$.
Specifically, for $p=1,3,5$, and $7$, we computed the average approximation ratio $r$ and the optimal probability $P$ for noise strengths $\eta \in \{0,0.001,0.002,\ldots,0.01\}$.
The simulations were performed using five DAX stocks—LIN.DE, BAYN.DE, VNA.DE, MTX.DE, and MUV2.DE—with investment constraint $B=2$.
The simulations were performed using the Qiskit DensityMatrix Simulator with 8192 measurement shots, and the COBYLA optimizer~\cite{Powell1994} was employed for parameter optimization.

\begin{figure}[h]
\centering
\includegraphics[width=1.0\textwidth]{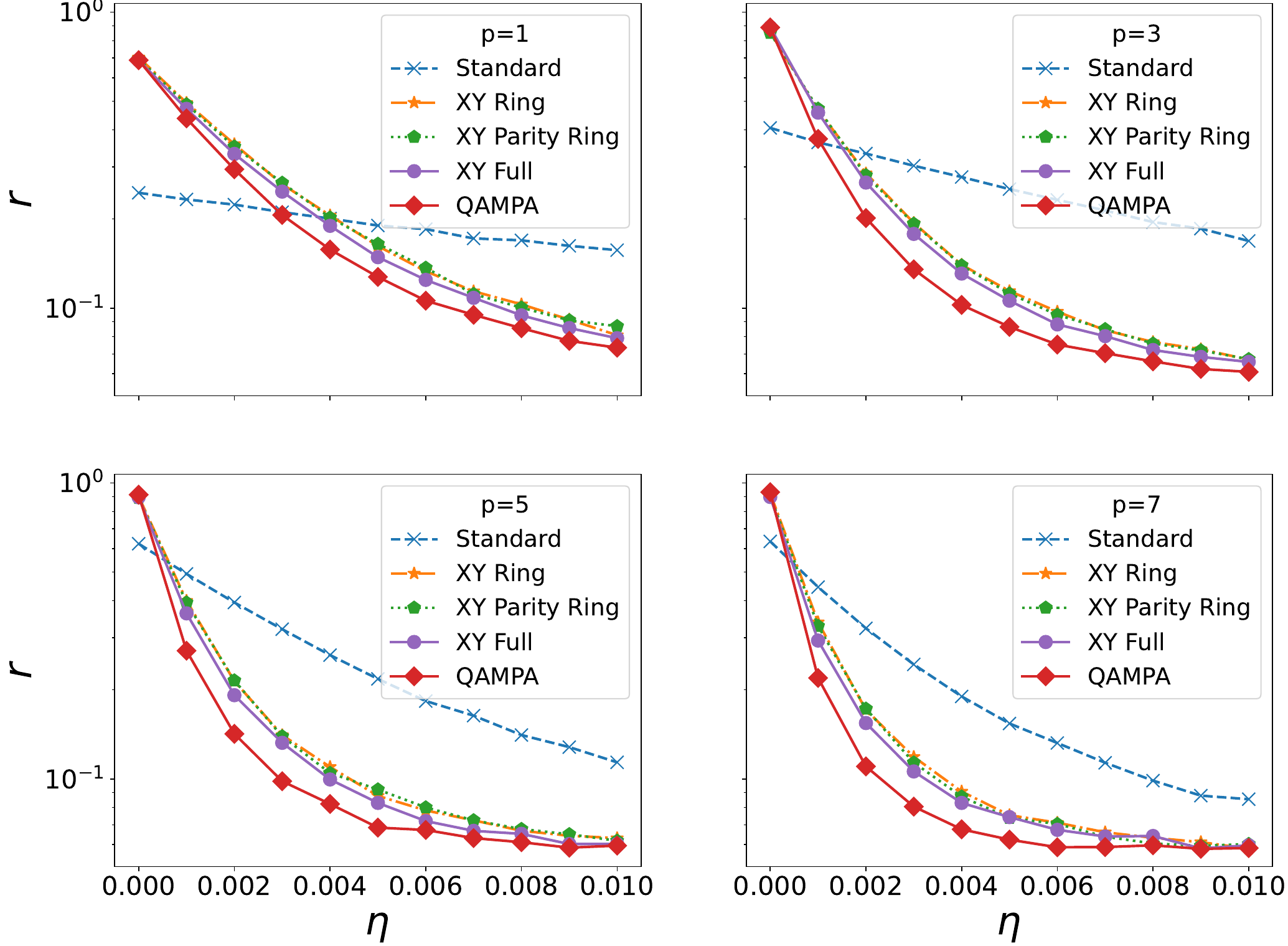}
\caption{Average approximation ratio as a function of depolarizing noise strength $\eta$ for five different Mixers, evaluated for $n=5$ assets with an investment constraint $B=2$ at QAOA depths $p=1,3,5$,and $7$.}\label{fig:noise_r}
\end{figure}

Figure~\ref{fig:noise_r} shows how the average approximation ratio $r$ varies with noise strength 
$\eta$ for different Mixers and QAOA depths $p$.
As $p$ increases, we observe that beyond a certain noise threshold, the Standard Mixer begins to outperform the XY Mixers. Specifically, the crossover noise strengths are summarized in Table~\ref{tab:crossover}.
Here, the crossover noise strengths are defined as the noise strengths at which the performance of the Standard Mixer and the XY Mixers are exchanged.

The observed crossover, in which the Standard Mixer outperforms the XY Mixers beyond a certain noise threshold, is attributed primarily to the significantly larger number of gates required by the XY Mixers, each contributing additional noise and leading to greater cumulative decoherence. As $p$ increases, circuit depth grows, widening the gate-count gap between the Standard Mixer and the XY Mixers and thereby amplifying this effect. Moreover, the XY Ring and XY Parity Ring Mixers exhibit relatively better noise tolerance than the XY Full Mixer and QAMPA due to their shallower circuit structures.

\begin{figure}[h]
\centering
\includegraphics[width=1.0\textwidth]{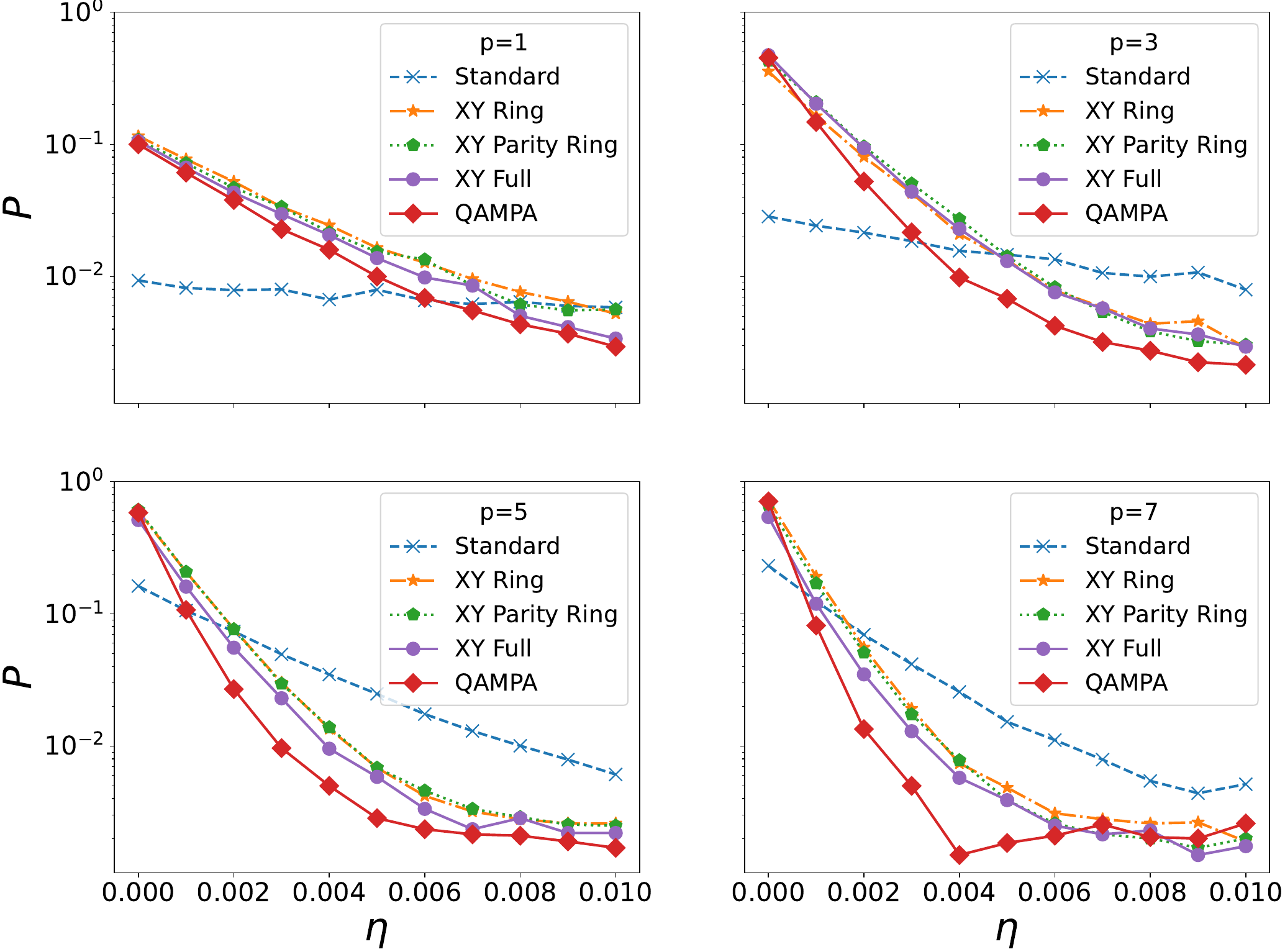}
\caption{Average approximation ratio as a function of depolarizing noise strength $\eta$ for five diﬀerent Mixers, evaluated for $n=5$ assets with an investment constraint $B=2$ at QAOA depths $p=1,3,5$, and $7$.}\label{fig:noise_P}
\end{figure}

Figure~\ref{fig:noise_P} shows the comparison of optimal solution probabilities $P$. Similar to the approximation ratio, we observe that as QAOA depth increases, the Standard Mixer becomes superior beyond certain noise strengths. The crossover occurs around  the values shown in Table~\ref{tab:crossover}.

The degradation of the XY Mixers in high-noise environments can be attributed to two structural characteristics inherent to their design. The first is the fragility of the initial state employed by the XY Mixers. Since this state relies on a carefully prepared superposition within the feasible subspace, it is highly susceptible to corruption under depolarizing noise. Once noise disturbs this structure, the algorithm can no longer exploit the intended constraints effectively.

\begin{figure}[h]
\centering
\includegraphics[width=1.0\textwidth]{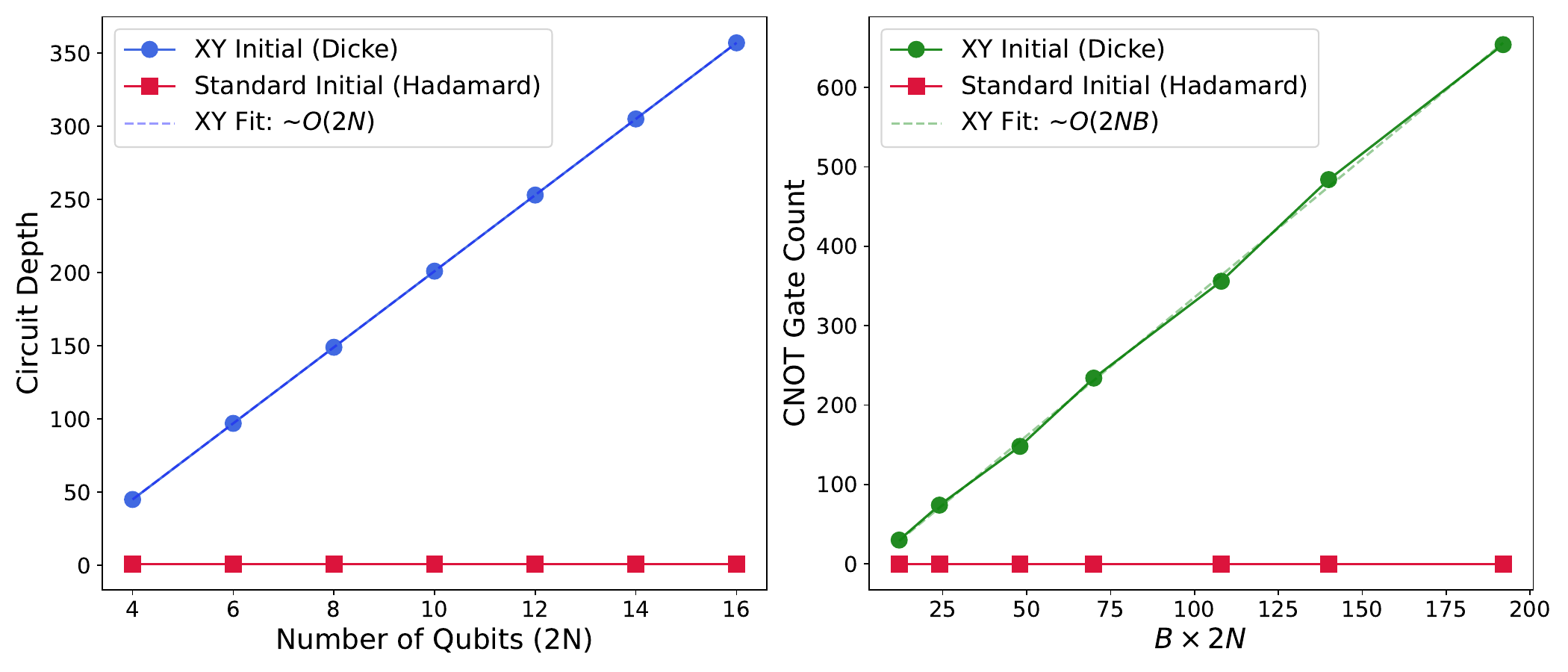}
\caption{Comparison of quantum circuit complexity for initial state preparation. The left panel shows the circuit depth as a function of the total number of qubits, and the right panel shows the CNOT gate count scaling against the product of qubits and Hamming weight ($B\times2N$). The Standard Mixer requires constant depth and no entanglement, whereas the Dicke state preparation for XY Mixers shows linear scaling in depth and significant entanglement cost, leading to greater susceptibility to decoherence.}
\label{fig:initial_state_comparison}
\end{figure}

The second factor is the overhead associated with preparing this initial state. Unlike the Standard Mixer, which requires only a single layer of Hadamard gates to generate the uniform superposition, the XY Mixers depend on a considerably larger number of two-qubit operations. 
To quantify this overhead, we compared the circuit complexity of the two initialization methods as a function of problem size. As shown in Fig.~\ref{fig:initial_state_comparison}, the Standard Mixer maintains a constant depth with zero CNOT gates regardless of system size. In contrast, the Dicke state preparation required for XY Mixers exhibits a linear increase in circuit depth and a substantial growth in the number of CNOT gates.For instance, with 10-qubits, the preparation circuit alone requires hundreds of gates.
These additional gates accumulate noise before the variational layers of QAOA are even applied, resulting in a significant loss of coherence. Consequently, the XY Mixers become less robust as noise increases, explaining their performance degradation compared with the Standard Mixer.

\begin{figure}[h]
\centering
\includegraphics[width=1.0\textwidth]{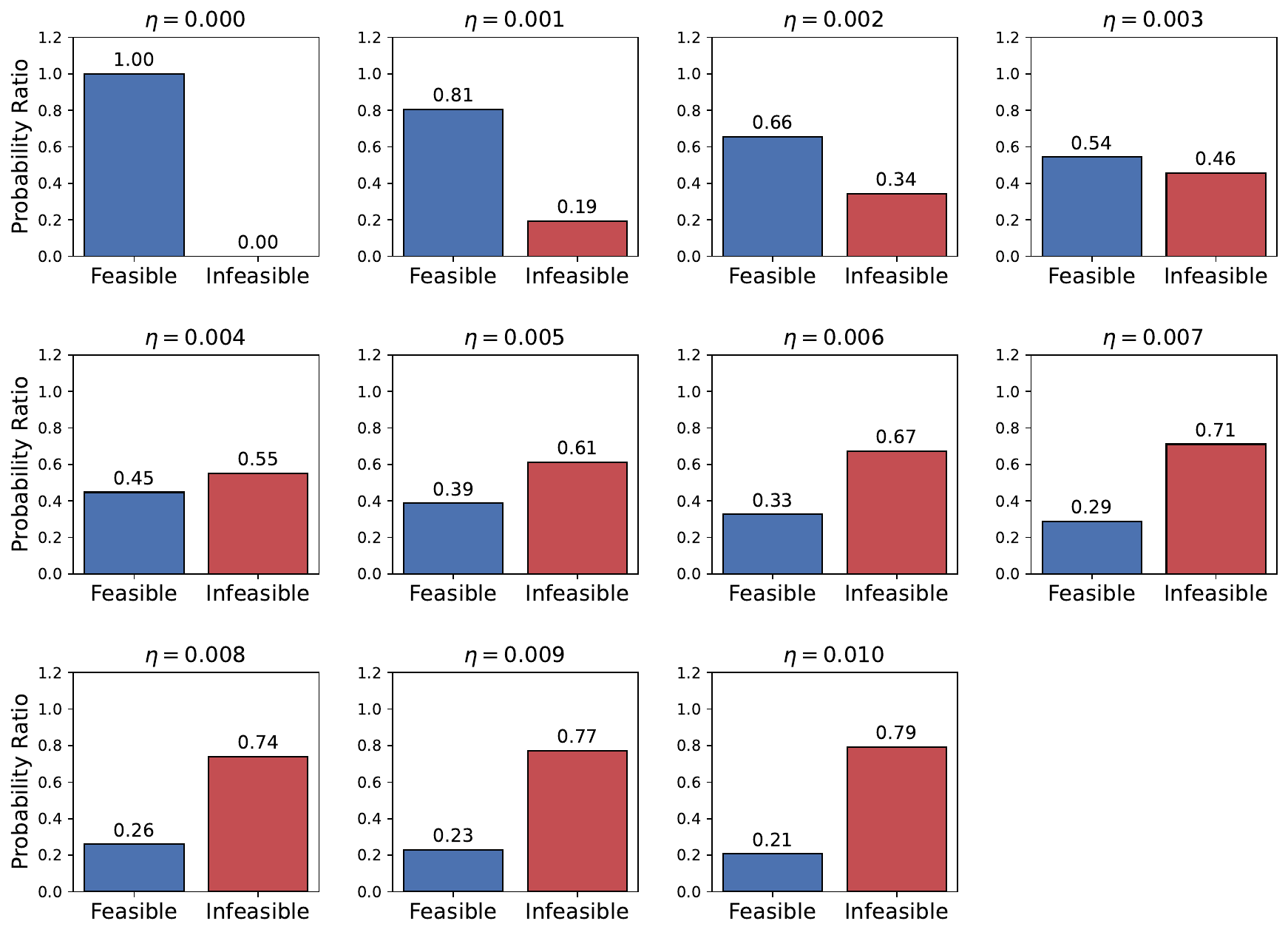}
\caption{Feasible-state probability $P_{\mathrm{feasible}}$ when preparing a Dicke state under depolarizing noise for $N=5$, $B=2$. 
Each data point is estimated from 20,000 shots. As the noise strength $\eta$ increases, $P_{\mathrm{feasible}}$ monotonically decreases, 
indicating that depolarizing noise drives the Dicke state out of the constraint-satisfying subspace.}
\label{fig:dicke_prob}
\end{figure}

To further elucidate the origin of the performance degradation of XY-based Mixers under noise, we examine how depolarizing noise affects the preparation of Dicke states, which constitute the initial states of these Mixers. Fig.~\ref{fig:dicke_prob} shows the probability $P_{\mathrm{feasible}}$ of measuring states that satisfy the constraint (feasible states), estimated from 20,000 shots for each noise strength $\eta$. As $\eta$ increases, the Dicke state component rapidly diminishes, indicating that depolarizing noise drives the system out of the subspace that satisfies the portfolio constraint. Since XY Mixers rely on the preservation of this subspace to restrict the dynamics to feasible solutions, the loss of Dicke state structure directly undermines their advantage. This observation provides a microscopic explanation for the crossover behavior observed in Figs.~\ref{fig:noise_r} and \ref{fig:noise_P}, where the Standard Mixer becomes superior beyond a certain noise threshold.

\begin{table}[h]
    \caption{Noise crossover thresholds for different QAOA depth $p$}
    \centering
    \begin{tabular}{ccc}
        \hline
        QAOA depth $p$ & $\eta$ in approximate ratio $r$ & $\eta$ in Optimal probability $P$\\
        \hline\hline
         1 & 0.003 & 0.005\\
         3 & 0.001 & 0.003\\
         5 & 0.0005 & 0.001\\
         7 & 0.0005 & 0.001\\
         \hline
    \end{tabular}
    \label{tab:crossover}
\end{table}

\section{Conclusion}

In this study, we extended the binary portfolio optimization problem considered in previous QAOA-based studies~\cite{Brandhofer2022} to a ternary formulation. To this end, we designed five types of Mixers—the Standard Mixer, XY Ring Mixer, XY Parity Ring Mixer, XY Full Mixer, and QAMPA—and conducted a comparative performance analysis. The simulation results showed that, in an ideal noiseless environment, the XY Mixers, particularly the XY Full Mixer and QAMPA, exhibited superior performance.
In contrast, when noise is taken into account, the XY Mixers demonstrate lower noise tolerance than the Standard Mixer, resulting in a substantial degradation in performance.
This degradation can be attributed to the large number of gates required for the initial-state preparation, as well as the intrinsic sensitivity of the initial state itself to noise.
Overall, these findings indicate that the choice of mixer critically depends on the noise level present in practical quantum hardware.

Compared to the binary formulation studied in previous QAOA-based work~\cite{Brandhofer2022}, extending the problem to a ternary representation effectively doubles the number of qubits required to encode each asset position, thereby enlarging the search space.
As a consequence, for a fixed number of assets, both the approximation ratio $r$ and Optimal probability $P$ are reduced in the noiseless setting.
Nevertheless, the qualitative behavior remains consistent with the binary case, in that the relative performance ordering among Mixers is preserved, with more expressive XY Mixers outperforming simpler ones in the absence of noise.
When noise is taken into account, a similar qualitative trend is observed in both binary and ternary formulations. However, an important distinction arises: while in the binary case XY mixers outperform the Standard Mixer already at $p=1$, in the ternary formulation the increased circuit depth and sensitivity to noise can lead to regimes in which the Standard Mixer achieves better performance even at $p=1$, depending on the noise strength.

From a practical perspective, our results indicate that shallow and low-overhead Mixers, such as the Standard or XY Ring Mixer, are preferable on near-term noisy devices, where the preparation of constraint-satisfying initial states is strongly affected by noise.
In contrast, more expressive Mixers like the XY Full Mixer and QAMPA become advantageous only in regimes where noise is sufficiently suppressed or mitigated, allowing the constraint structure to be reliably preserved.
These findings highlight the importance of selecting the mixer and ansatz at the outset based on the noise characteristics of the hardware, and motivate future efforts to design Mixers and initial states that remain effective even when strict constraint preservation cannot be guaranteed.

\vspace{0.5cm}
\noindent\textbf{Acknowledgements} The numerical calculations were performed on computing resources aquired through collavorative research funds from KDDI Research, Inc.

\vspace{0.5cm}
\noindent\textbf{Author Contributions} S.Y. contributed to the implementation of the software and the writing of the manuscript draft. S.W. contributed to the mathematical design of the Mixers (Sec.~\ref{subsec:Mixer_unitary}) and the method for Dicke state preparation (Appendix~\ref{appendix:dicke}). S.Y. and S.W. contributed equally to this work. M.K. contributed to the proposal of the methodology, discussions, and revision of the manuscript. K.S. contributed to the proposal of the methodology and discussions. T.N. contributed to the proposal of the methodology, discussions, and revision of the manuscript.

\vspace{0.5cm}
\noindent\textbf{Funding} This work was supported by JSPS KAKENHI under Grant No. JP25K00215 (M.K.) and JST ASPIRE under Grant No. JPMJAP24C2 (M.K.)

\vspace{0.5cm}
\noindent\textbf{Data Availability} The source data supporting the findings of this study are available in the GitHub repository.
The simulation data for the ternary portfolio optimization with $N=5$ and $8$, as well as the data obtained under depolarizing noise noise, are available at: \url{https://github.com/Shintaro-Yamamura/QAOA-Ternary-Portfolio}.

\section*{Declarations}
\noindent\textbf{Conflict of interest} The authors declare no conflict of interest.

\appendix

\section{Estimate return and risk in a portfolio}\label{appendix:Estimate return and risk in a portfolio}

\subsection{Data Acquisition}
To construct the portfolio optimization instances, we utilized historical market data for the constituents of the German stock index (DAX 30).
The daily adjusted closing prices for the assets were retrieved from Yahoo Finance using the open-source Python library \texttt{yfinance}~\cite{yfinance}.
The dataset covers a period of 5 years, from 2016/01/01 to 2020/12/31.
The assets used in the simulations were selected from this dataset.
Missing data points, if any, were handled by dropping the corresponding dates.
Based on these historical prices, the daily simple returns $r_{t,i}$ were calculated, which served as the input for estimating the expected returns $\mu_i\;(i=1,2,\ldots,N)$ and the covariance matrix $\sigma_{ij}\;(i,j=1,2,\ldots,N)$ described in the following subsections.

\subsection{Estimate expected returns}

Given the daily simple returns $\{r_{t,i}\}_{t=1,2,\dots,T}^{i=1,2,\dots,N}$ of each asset, we estimate the annualized expected return of asset $i$ from its sample mean
daily return. Let
\begin{equation}
\bar{r}_i = \frac{1}{T}\sum_{t=1}^{T} r_{t,i},\label{eq:bar_return}
\end{equation}
denote the empirical average daily return. The annualized expected return $\mu_i$ is then
computed using geometric compounding under the assumption of 252 trading days per year, i.e.,
\begin{equation}
    \mu_i = \left[\prod_{t=1}^{T}(1+r_{t,i})\right]^{\frac{252}{T}} - 1.
    \label{eq:annual_return}
\end{equation}
This corresponds to the standard transformation from daily to annual returns in
portfolio theory~\cite{Markowitz1952,Campbell1997,Cochrane2009,Elton2014}. 

\subsection{Estimate covariance matrix}\label{subsec:Estimate_cov}
Given the matrix of daily simple returns $\{r_{t,i}\}_{t=1,2,\dots,T}^{i=1,2,\dots,N}$, we estimate
the daily return covariance matrix using the usual unbiased sample covariance estimator:
\begin{equation}
  {\sigma}^{(\mathrm{daily})}_{ij}
  = \frac{1}{T - 1} \sum_{t=1}^{T}
    \bigl(r_{t,i} - \bar{r}_i\bigr)
    \bigl(r_{t,j} - \bar{r}_j\bigr).
  \label{eq:daily_cov}
\end{equation}
To obtain an annualized covariance matrix, we multiply the daily covariance estimates by 252:
\begin{equation}
  \sigma_{ij} = 252 \times \sigma^{(\mathrm{daily})}_{ij}.
  \label{eq:annual_cov}
\end{equation}

\section{How to create the initial state}\label{appendix:dicke}

The initial state represented by Eq.~\eqref{eq:XY_initial} can be identified with a Dicke state~\cite{Dicke1954}.
In this appendix, we describe the preparation of the initial state used in this work, based on the Dicke state construction.

\subsection{Fixed-Hamming-weight subspace and feasibility mapping}

As formulated in Sec.~\ref{sec2}, each asset $i=1,\ldots,N$ is represented by a ternary variable
\begin{equation}
    z_i \in \{+1,0,-1\},
\end{equation}
corresponding to a long position, no-position, and a short position, respectively.
Using the two-qubit encoding introduced in Sec.~\ref{subsec:encording and cost hamiltonian}, we assign binary variables $x_i^+,x_i^-\in\{0,1\}$ such that
\begin{equation}
    z_i = x_i^+ - x_i^-.
\end{equation}
A portfolio configuration $(z_1,\ldots,z_N)$ is therefore represented by a $2N$-qubit computational basis state
\begin{equation}
    |x_1^-,x_1^+,\ldots,x_N^-,x_N^+\rangle.
\end{equation}
The investment constraint considered in this work imposes
\begin{equation}
    \sum_{i=1}^{N} z_i = B.
    \label{eq:appendix_const}
\end{equation}

To convert this constraint into a fixed-Hamming-weight condition, we introduce the complement variable
\begin{equation}
    \tilde{s}_i = 1 - x_i^- .
\end{equation}
This is merely a reparameterization of the short qubit and does not introduce an additional degree of freedom; in the actual circuit it is implemented by applying Pauli-$\hat{X}$ gates to all short qubits after the Dicke state has been prepared.

Using $\tilde{s}_i$, the constraint can be rewritten as
\begin{align}
    \sum_{i=1}^{N} z_i
    &= \sum_{i=1}^{N}\left(x_i^+ - x_i^-\right)
     = \sum_{i=1}^{N}\left(x_i^+ + \tilde{s}_i\right) - N=B,
\end{align}
and therefore Eq.~\eqref{eq:appendix_const} is equivalent to
\begin{equation}
    \sum_{i=1}^{N}\left(x_i^+ + \tilde{s}_i\right) = N + B.
\end{equation}
That is, the feasible subspace specified by the portfolio constraint corresponds exactly to the set of bitstrings of length $2N$ whose Hamming weight is fixed to
\begin{equation}
    k = N + B.
\end{equation}
Consequently, preparing a Dicke state $|D_{N+B}^{2N}\rangle$ on the variables $(x_i^+,\tilde{s}_i)$ directly yields a quantum state supported only on feasible portfolios. This provides an initial state entirely restricted to the feasible domain, suitable for XY Mixers.

\subsection{Definition of Dicke state}
The Dicke state of Hamming weight $k$ on $n$ qubits is defined by
\begin{equation}
    |D_k^n\rangle
    = \frac{1}{\sqrt{\binom{n}{k}}}
      \sum_{x\in\{0,1\}^n,\ \mathrm{wt}(x)=k} |x\rangle .
\end{equation}
For our ternary portfolio formulation, the initial state required by the XY Mixers, Eq.~\eqref{eq:XY_initial}, can thus be written as an $\hat{X}$-transformed Dicke state.
Let $\hat{X}_i^{-}$ denote the Pauli-$\hat{X}$ operator acting on the short qubit of asset $i$. The initial state used in the circuit is
\begin{equation}
    |\psi_0\rangle_{M_{XY}}
    = \bigotimes_{i=1}^N \hat{X}_i^{-} |D_{N+B}^{2N}\rangle .
\end{equation}
Equivalently, in terms of the reparameterization $(x_i^+,\tilde{s}_i)$ introduced above, this state can be viewed as the Dicke state $|D_{N+B}^{2N}\rangle$ on a fixed-Hamming-weight subspace, and therefore the algorithm evolves within the feasible subspace throughout the QAOA layers. This is in contrast to the Standard Mixer, which explores the full Hilbert space.

\subsection{Deterministic preparation method}\label{subsec:Deterministic preparation method}

A deterministic circuit for preparing Dicke states with no ancilla qubits and with gate complexity $\mathcal{O}(kn)$ and depth $\mathcal{O}(n)$ is described in \cite{Bartschi2019}.
The method is based on the recursive decomposition
\begin{equation}
    |D_{\ell}^{n}\rangle
    = \sqrt{\frac{\ell}{n}}\, |D_{\ell-1}^{n-1}\rangle\otimes |1\rangle
      + \sqrt{\frac{n-\ell}{n}}\, |D_{\ell}^{n-1}\rangle\otimes |0\rangle,
    \label{eq:appendix_D}
\end{equation}
which expresses a Dicke state of size $n$ and Hamming weight $\ell$ in terms of two Dicke states of size $n-1$.
By implementing a Split-and-Cyclic-Shift (SCS) operation that generates the amplitude ratios in Eq.~\eqref{eq:appendix_D}, a Dicke state $|D_{k}^{n}\rangle$ can be constructed inductively from the computational basis state
\begin{equation}
    |0\rangle^{\otimes (n-k)} |1\rangle^{\otimes k}.
\end{equation}
Following the method presented in \cite{Bartschi2019}, we first generate the Dicke state $|D_{N+B}^{2N}\rangle$ and subsequently apply Pauli-$X$ gates to all short qubits, as described above. Since the resulting state lies entirely within the feasible Hamming-weight subspace, the QAOA evolution with XY-based Mixers preserves the constraint $\sum_{i} z_i = B$ without requiring penalty terms.

\section{Quantum Circuits}\label{appendix:Quantum Circuits}
In this section, we describe the construction of quantum circuits for the cost unitary, the mixer unitary, and the initial state in XY Mixers.

\subsection{Cost unitary}
\begin{figure}[h]
  \centering
  \resizebox{1.0\linewidth}{!}{%
  \begin{quantikz}
    \lstick{$|x_i^-\rangle$} & \ctrl{1}\gategroup[2,steps=3,style={dashed,rounded
    corners,fill=blue!20, inner
    xsep=2pt},background,label style={label
    position=above,anchor=south,yshift=0.0cm}]{$e^{-i\gamma W_{ij}\hat{Z}_i^-\hat{Z}_j^-}$} & & \ctrl{1} & \ctrl{3}\gategroup[4,steps=3,style={dashed,rounded
    corners,fill=blue!20, inner
    xsep=2pt},background,label style={label
    position=above,anchor=south,yshift=0.0cm}]{$e^{-i\gamma W_{ij}\hat{Z}_i^-\hat{Z}_j^-}$} & & \ctrl{3} &  & & & & & &\\
    \lstick{$|x_i^+\rangle$} & \targ{} & \gate{R_z\!\left(2\gamma W_{ij}\right)} & \targ{} &  &  &  & \ctrl{1}\gategroup[2,steps=3,style={dashed,rounded
    corners,fill=blue!20, inner
    xsep=2pt},background,label style={label
    position=above,anchor=south,yshift=0.0cm}]{$e^{-i\gamma W_{ij}\hat{Z}_i^+\hat{Z}_j^-}$} & & \ctrl{1} & \ctrl{2}\gategroup[3,steps=3,style={dashed,rounded
    corners,fill=blue!20, inner
    xsep=2pt},background,label style={label
    position=above,anchor=south,yshift=0.0cm}]{$e^{-i\gamma W_{ij}\hat{Z}_i^-\hat{Z}_j^+}$} & & \ctrl{2} &\\
    \lstick{$|x_j^-\rangle$} & & & & & & & \targ{} & \gate{R_z\!\left(-2\gamma W_{ij}\right)} & \targ{} & & & &\\
    \lstick{$|x_j^+\rangle$} & & & & \targ{} & \gate{R_z\!\left(-2\gamma W_{ij}\right)} & \targ{} &  & & & \targ{} & \gate{R_z\!\left(2\gamma W_{ij}\right)} & \targ{} &
  \end{quantikz}
  }
  \caption{Circuit implementation of the two-qubit component in the cost unitary $\hat{U}_F(\gamma)$, decomposed into CNOT and $\hat{R}_z$ gates acting independently on short and long qubits.}
  \label{fig:cost_unitary_circuit}
\end{figure}
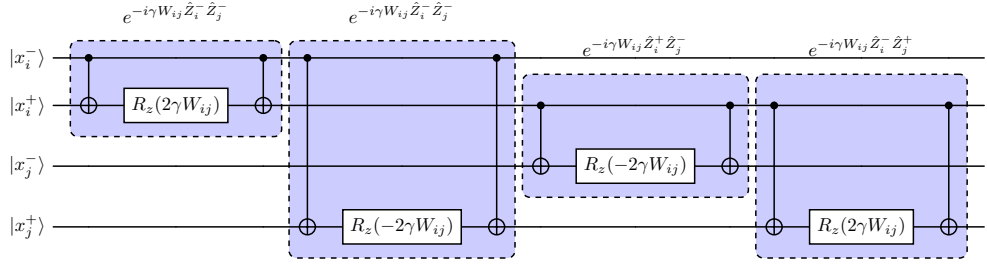

The cost unitary $\hat{U}_F(\beta)$ is given by Eq.~\eqref{eq:cost_hamiltonian} and consists of a single-qubit gate part and two-qubit gate part. 
The single-qubit part can be implemented using $\hat{R}_z$ rotations, while the two-qubit part can be implemented as shown in the Fig~\ref{fig:cost_unitary_circuit}.

\subsection{Mixer unitary}
\begin{figure}[h]
  \centering
  \resizebox{1.0\linewidth}{!}{%
  \begin{quantikz}
    \lstick{$|x_i^-\rangle$} & \gate{R_x\!\left(-\pi/2\right)}\gategroup[3,steps=5,style={dashed,rounded
    corners,fill=blue!20, inner
    xsep=2pt},background,label style={label
    position=above,anchor=south,yshift=0.0cm}]{$\hat{R}_{x_i^-,x_j^-}^{(XY)}(\beta)$} & \ctrl{2} & \gate{R_x\!\left(-2\beta \right)} & \ctrl{2}& \gate{R_x\!\left(\pi/2\right)} & & & & & &\\
    \lstick{$|x_i^+\rangle$} & & & & & & \gate{R_x\!\left(-\pi/2\right)}\gategroup[3,steps=5,style={dashed,rounded
    corners,fill=blue!20, inner
    xsep=2pt},background,label style={label
    position=below,anchor=north,yshift=-0.2cm}]{$\hat{R}_{x_i^+,x_j^+}^{(XY)}(\beta)$} & \ctrl{2} & \gate{R_x\!\left(-2\beta \right)} & \ctrl{2} & \gate{R_x\!\left(\pi/2\right)} &\\
    \lstick{$|x_j^-\rangle$} & \gate{R_x\!\left(\pi/2\right)} & \targ{} & \gate{R_z\!\left(2\beta\right)} & \targ{} & \gate{R_x\!\left(-\pi/2\right)} & & & & & &\\
    \lstick{$|x_i^+\rangle$} & & & & & & \gate{R_x\!\left(\pi/2\right)} & \targ{} & \gate{R_z\!\left(2\beta \right)} & \targ{} & \gate{R_x\!\left(\pi/2\right)} &
  \end{quantikz}
  }
  \caption{Circuit implementation of the two-qubit component in the XY Mixer unitary $\hat{U}_M(\beta)$, decomposed into $\hat{R}_x$ and $\hat{R}z$ rotations acting independently on the short and long qubits.
}
  \label{fig:xy_unitary_circuit}
\end{figure}
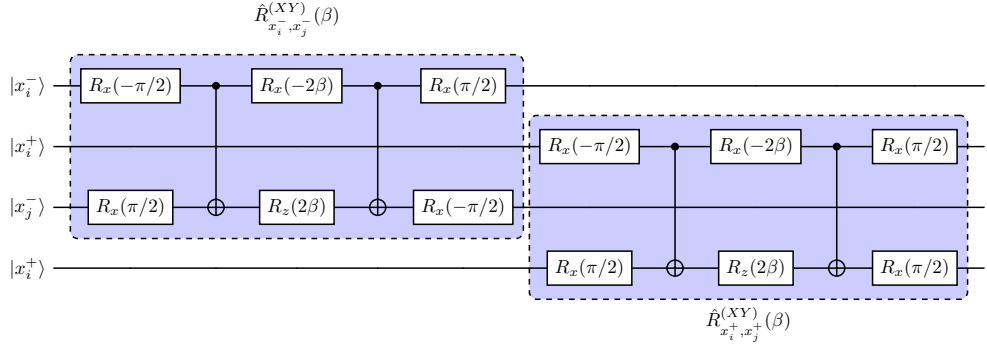

The implementation of the XY mixer is described below.
The $\hat{R}_{i,j}^{(XY)}(\beta)$ operator, acting independently on the long and short qubits, is implemented as shown in Fig.~\ref{fig:xy_unitary_circuit}.

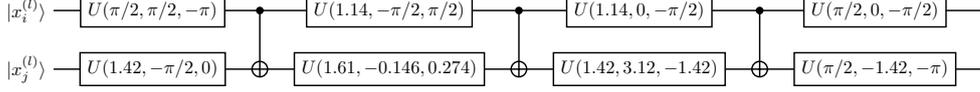
\begin{figure}[h]
  \centering
  \resizebox{1.0\linewidth}{!}{%
  \begin{quantikz}
    \lstick{$|x_i^{(l)}\rangle$} & \gate{U(\pi/2,\pi/2,-\pi)} & \ctrl{1} & \gate{U(1.14,-\pi/2,\pi/2)} & \ctrl{1} & \gate{U(1.14,0,-\pi/2)} & \ctrl{1} & \gate{U(\pi/2,0,-\pi/2)} &\\
    \lstick{$|x_j^{(l)}\rangle$} & \gate{U(1.42,-\pi/2,0)} & \targ{} & \gate{U(1.61,-0.146,0.274)} & \targ{} & \gate{U(1.42,3.12,-1.42)} & \targ{} & \gate{U(\pi/2,-1.42,-\pi)} &
  \end{quantikz}
  }
  \caption{Implementation of the two-qubit QAMPA block obtained by synthesizing the unitary Eq.~\eqref{eq:QAMAP_2} with fixed parameters $\beta=\gamma=1.0$.
  The unitary is synthesized as a three-CNOT entangling block with single-qubit rotations $\hat{U}(\theta,\phi,\lambda)=\hat{R}_z(\phi)\hat{R}_y(\theta)\hat{R}_z(\lambda)$, yielding the corresponding QAMPA two-qubit component.}
  \label{fig:qampa_circuit}
\end{figure}

QAMPA is defined in Eq.~\eqref{eq:QAMPA} , and its concrete implementation proceeds as follows.
Here we describe the implementation of the exponential term
\begin{equation}
    e^{\, i\beta(\hat{X}_i^{(\ell)}\hat{X}_j^{(\ell)}
               +\hat{Y}_i^{(\ell)}\hat{Y}_j^{(\ell)})
     - i\gamma W_{ij}\hat{Z}_i^{(\ell)}\hat{Z}_j^{(\ell)}},\label{eq:QAMAP_2}
\end{equation}
while the remaining terms can be implemented in an analogous manner.
In practice, we employ a KAK (Cartan) decomposition~\cite{Tucci2005, Zhang2003, Shende2006} to synthesize the corresponding two-qubit unitary into a standard block composed of single-qubit generalized rotations
$\hat{U}(\theta,\phi,\lambda)$ and CNOT gates.
The decomposition algorithm optimizes the synthesis so as to minimize the number of CNOT gates, yielding a block that typically consists of three CNOT gates interleaved with local single-qubit $\hat{U}$ gates. This achieves an implementation that is equivalent to the target unitary up to a global phase, while using the minimal entangling resources.
The resulting circuit structure is shown in Fig.~\ref{fig:qampa_circuit}.

\subsection{Initial state}
\begin{figure}[h]
\centering
  \resizebox{1.0\linewidth}{!}{%
  \begin{quantikz}
    \lstick{$|x_1^{-}=0\rangle$}  & \slice[style=red, label style={pos=1, anchor=north}]{$|0\rangle^{\otimes N-B}|1\rangle^{\otimes N+B}$}& & & \gate[5]{SCS_{5,4}} & \gate[4]{SCS_{4,3}} & \gate[3]{SCS_{3,2}} & \gate[2]{SCS_{2,1}} & \slice[style=red, label style={pos=1, anchor=north}]{$|D_{N+B}^{2N}\rangle$} & \gate{X} & \rstick[wires = 6]{$|\psi_0\rangle_{M_{XY}}$}\\
    \lstick{$|x_1^{+}=0\rangle$}  & & & \gate[5]{SCS_{6,4}} & & & & & & &\\
    \lstick{$|x_2^{-}=0\rangle$}  & \gate{X} & & & & & & & & \gate{X} &\\
    \lstick{$|x_2^{+}=0\rangle$}  & \gate{X} & & & & & & & & &\\
    \lstick{$|x_3^{-}=0\rangle$}  & \gate{X} & & & & & & & & \gate{X} &\\
    \lstick{$|x_3^{+}=0\rangle$}  & \gate{X} & & & & & & & & &
  \end{quantikz}
  }
\caption{Quantum circuit for the preparation of the initial state $|\psi_0\rangle_{M_{XY}}$ with $N=3,B=1$. The circuit is constracted following the SCS operation.} 
\label{fig:dicke_state}
\end{figure}
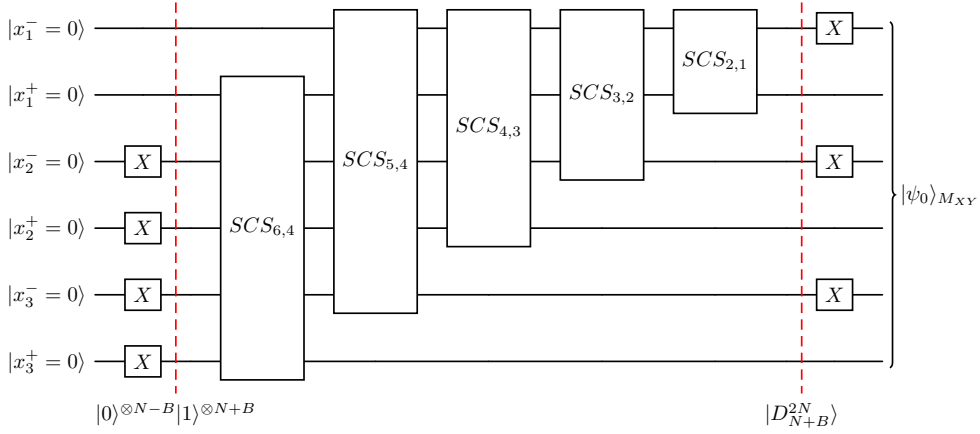

The initial state can be prepared according to~\ref{subsec:Deterministic preparation method}.
Figure~\ref{fig:dicke_state} shows the circuit for $N=3$ and $B=1$.

\end{document}